\documentclass[apj]{emulateapj}
\usepackage{graphicx}
\usepackage[dvips]{color}
\usepackage[utf8]{inputenc}
\usepackage{natbib}
\usepackage{amsmath,amssymb}

\def\lsim{\raise0.3ex\hbox{$\;<$\kern-0.75em\raise-1.1ex\hbox{$\sim\;$}}}
\def\gsim{\raise0.3ex\hbox{$\;>$\kern-0.75em\raise-1.1ex\hbox{$\sim\;$}}}

\def\be{\begin{equation}}
\def\ee{\end{equation}}

\def\theta{\vartheta}

\renewcommand{\vec}[1]{\boldsymbol{#1}}

\shortauthors{Andersen et al.}
\shorttitle{High-energy neutrinos from Galactic superbubbles}

\begin{document}

\author{K.~J.~Andersen$^{1}$, M.~Kachelrie\ss$^{1}$, D.~V.~Semikoz$^{2,3}$}

\affil{
$^1$Institutt for fysikk, NTNU, Trondheim, Norway\\
$^{2}$APC, Universite Paris Diderot, CNRS/IN2P3, CEA/IRFU,
Observatoire de Paris, Sorbonne Paris Cite, 119 75205 Paris, France\\
$^{3}$National Research Nuclear University MEPHI
(Moscow Engineering Physics Institute), Kashirskoe highway 31,
115409 Moscow, Russia}

\title{High-energy neutrinos from Galactic superbubbles}

\begin{abstract}
We study the propagation of cosmic rays generated by sources residing inside
superbubbles. We show that the enhanced magnetic field in the bubble wall
leads to an increase of the interior cosmic ray density. Because of the large
matter density in the wall, the probability for cosmic ray interactions on gas
peaks there. As a result, the walls of superbubbles located near young
cosmic ray sources emit efficiently neutrinos. We apply this scenario
to the Loop~I and Local Superbubble: These bubbles are sufficiently near
such that cosmic rays from a young source as Vela interacting in the
bubble wall can generate a substantial fraction of the observed astrophysical
high-energy neutrino flux below $\sim {\rm few} \times 100$\,TeV.
\end{abstract}

\keywords{High energy cosmic rays, high-energy neutrinos and photons}


\maketitle

\section{Introduction}
%
High-energy astrophysical neutrinos are a key tool 
to understand the non-thermal
universe~\citep{Gaisser:1994yf}. They are produced together with photons
in interactions of cosmic rays (CR) on matter and background photons close
to their sources and during propagation. These neutrinos travel undisturbed,
being neither absorbed as high-energy photons nor deflected in
magnetic fields as charged particles. These two properties distinguish
neutrinos as a unique tracer of CR sources.

The discovery of astrophysical neutrinos in 2013 by the IceCube 
collaboration marked the beginning of neutrino 
astronomy~\citep{Aartsen:2013jdh,Aartsen:2014gkd}.
There are two main experimental channels to detect such neutrinos. Using the
tracks of muons produced in interactions of muon neutrinos one can measure
the neutrino arrival directions very precisely, while its energy can be
only estimated within a factor of a few~\citep{Aartsen:2016xlq}. To avoid the
atmospheric neutrino background, the energy spectrum of astrophysical neutrinos
in this channel is measured above 200\,TeV. Because neutrinos of this energy
are heavily absorbed in the Earth, the spectrum is dominated by a thin strip
around the horizon. It is consistent with a $1/E^{2.1}$ power law which is
predicted in many models of extragalactic neutrino sources~\citep{Stecker:1991vm,Mannheim:1995mm,Waxman:1997ti,Loeb:2006tw}.

In a second channel using cascade events inside the IceCube detector one
can detect electron neutrinos interacting via the charge current and
additionally all neutrino flavours interacting via neutral currents. In this
channel the energy of a neutrino can be measured with up to 10\% accuracy 
but its derived arrival direction has an error which is typically larger
than $10^\circ$. The energy spectrum of astrophysical neutrinos derived
in this channel is close to $1/E^{2.5}$~\citep{Aartsen:2017mau}. Such a
steep spectrum challenges an extragalactic origin of this component, since
the accompanying photons would overshoot the bounds on the diffuse background
of extragalactic gamma-rays~\citep{Berezinsky:2010xa,Murase:2013rfa}. Moreover,
the all-sky spectrum in this channel is consistent with a continuation of the 
all-sky spectrum in gamma-rays measured by Fermi-LAT~\citep{Neronov:2014uma}.  

Gamma-rays observed at the highest energies, i.e.\ in the TeV range, are
dominated by the Galactic contribution. \citet{Neronov:2014uma}
suggested that the four-year dataset of IceCube~\citep{Aartsen:2015knd} shows
at the highest energies $E>100$\,TeV in the cascade channel evidence for a
Galactic component. Two-component models with galactic and 
extragalactic contributions were suggested by 
\citet{Neronov:2016bnp} and \citet{Palladino:2017qda} to explain the
data in both channels. A non-zero Galactic contribution was obtained
also more recently in a multi-component fit performed in 
Ref.~\citep{Palladino:2018evm}. Finally, ~\citet{Neronov:2018ibl}
uncovered the electromagnetic counterpart of the IceCube signal using
Fermi-LAT data in the multi-TeV energy range. Since at these energies photons
are strongly attenuated by pair-production on cosmic background photons, 
the detection of a TeV photon counterpart demonstrates that this signal has 
a largely Galactic origin.

Note that the Galactic component may consist of two contributions: A component
from the Galactic plane and a local contribution at higher Galactic latitudes,
$10^{\circ} <b < 50^{\circ}$~\citep{Neronov:2015osa}. The first one contains the
``guaranteed'' contribution from diffuse Galactic CRs scattering on gas
in the Galactic plane, which is however both too small and too concentrated at 
small latitudes, $|b|\lsim 1^\circ$, to explain the IceCube
observations~\citep{Berezinsky:1992wr,Evoli:2007iy,Kachelriess:2014oma}. 
Additionally, the contribution from the Galactic
plane was restricted by ANTARES measurements~\citep{Albert:2017oba} 
and more recently
by IceCube~\citep{Aartsen:2017ujz}. Taken at face value, these measurements
seem to require an extragalactic origin of these astrophysical neutrinos.
On the other hand, the  most recent six-year cascade data of IceCube are still
consistent with a soft $1/E^{2.5}$ energy spectrum~\citep{Aartsen:2017mau}.
Last but not least, the TeV photon counterpart detected by 
\citet{Neronov:2018ibl} requires extended neutrino emission at large Galactic
latitudes.

In this {\em Letter\/}, we suggest as resolution of this puzzle that the soft
$1/E^{2.5}$ component in the astrophysical neutrino intensity is produced
locally by CRs interacting in the walls of nearby superbubbles. We investigate
as a possible realisation of this scenario that CRs interact in the walls
of the Local Bubble and the Loop~I superbubble, in particular in the part
which forms the interface between these two superbubbles. The large angular
extension of Loop~I on the sky explains why
the arrival directions of this Galactic component are not concentrated in
the Galactic disk. Moreover, the small distance to Loop~I is the
reason why a single source can dominate the Galactic neutrino flux.
The required CR flux may be delivered by a recent young
nearby supernova, such as Vela, close to or inside Loop~I.

This work is organised as follows: We first
examine the propagation of CRs emitted by a bursting
source inside an idealised (super-) bubble. Then we discuss the case
of the Loop~I and the local superbubble. Combining
our previous findings, we calculate the secondary
fluxes produced by CR interactions in the wall of Loop~I
and show that the
resulting neutrino flux can be a substantial fraction
of the observed one below $\sim {\rm few} \times 100$\,TeV.

\section{CR propagation in idealised (super-) bubbles}
%
Massive stars lose a significant fraction of their mass  in the form
of a stellar wind. As this wind expands, it collides with the  gas in the
interstellar medium (ISM), creating a low density bubble which expands over
time. Once the star explodes at the end of its fusion cycle as a core-collapse
supernova, a shock wave is injected into the ISM. The shock expands quickly
until it reaches the bubble wall where it is typically stopped~\citep{vanMarle:2015oca}.
Since massive stars are formed in clusters, the wind-blown bubbles of the
individual stars
encounter each other as they expand and merge to form a single
superbubble~\citep{vanMarle:2012bs,Krause:2013qar}. The shape of these superbubbles is determined
in particular by the pre-existing density inhomogeneities and magnetic fields,
as well as the positions of the first SNe. Simulations~\citep{2000A&A...361..303B,Schulreich:2017dyt} show
that the bubble walls are fragmented and twisted, and outflows away from
the Galactic plane may open up the bubble~\citep{2000A&A...361..303B}.

In view of this intricate geometry, we idealise the bubble in our numerical
simulations as follows: We assume for the magnetic field
$\vec B(\vec x) $ and density $n(\vec x)$ profiles perpendicular to the
Galactic plane $(x,y)$ a cylindrical symmetry, i.e.\ we imply that the changes as function of the Galactic height $z$ are small compared on the considered length scales. Then  $\vec B(\vec x) $ and $n(\vec x)$ are only functions of $r=\sqrt{x^2+y^2}$.
Inspired by~\citet{vanMarle:2015oca} we set the 
magnetic field strength inside the
bubble to $B_{\rm tot}=0.1\mu$G, $12\mu$G in the wall and $5\mu$G outside.
We assume that the energy density in the regular and turbulent field are
equal. For the later, we distribute field modes  between $L_{\max}=10$\,pc and $L_{\min}=10$\,AU according to an isotropic Kolmogorov power spectrum. In the calculation of CR trajectories, we include modes down to $\min_{\vec x}[R_L(\vec B(\vec x))]/3$, i.e.\ down to one third the smallest Larmor radius $R_L=cp/(eB)$ considered. For the generation of the turbulent field modes, we use the algorithm proposed in Ref.~\citep{1999ApJ...520..204G}.  We model the regular magnetic field outside the bubble in the Galactic disc to be constant and pointing in one direction, i.e.\ we neglect the curvature of the spiral arms on the scales we considered.
For the direction of the regular field inside the bubble,
we use a clockwise field for $y>0$ and an anticlockwise for $y<0$. 
Such a configuration corresponds to the naive picture that a uniform field 
is driven by a central explosion towards the South and North side of the 
bubble which is supported by analytical and numerical 
arguments~\citep{vanMarle:2015oca} 

We choose the radius $R$ of the bubble as $R=50$\,pc and set the width of the
bubble wall to 2\,pc, i.e.\ the wall extends between 49 and 51\,pc.
In order to have a smooth transition, we use the function
$T(r)=\{1+\exp[-(r-r_0)/w_0]\}^{-1}$ with $w_0=0.1$\,pc to interpolate between
different regions.
Then we inject CRs at the center of the bubble, calculate
their trajectories using the Lorentz equation and record the resulting surface
density $n={\rm d}N/{\rm d}\Omega={\rm d}N/(2\pi{\rm d}r)$. We consider
two cases for the injection history: A continuous source and a bursting  source.
In the latter case, we  record the CR density $n$ after
 $t=\{300,1000,3000,10000\}$\,yr.

\begin{figure*}
\includegraphics[width=\columnwidth,angle=0]{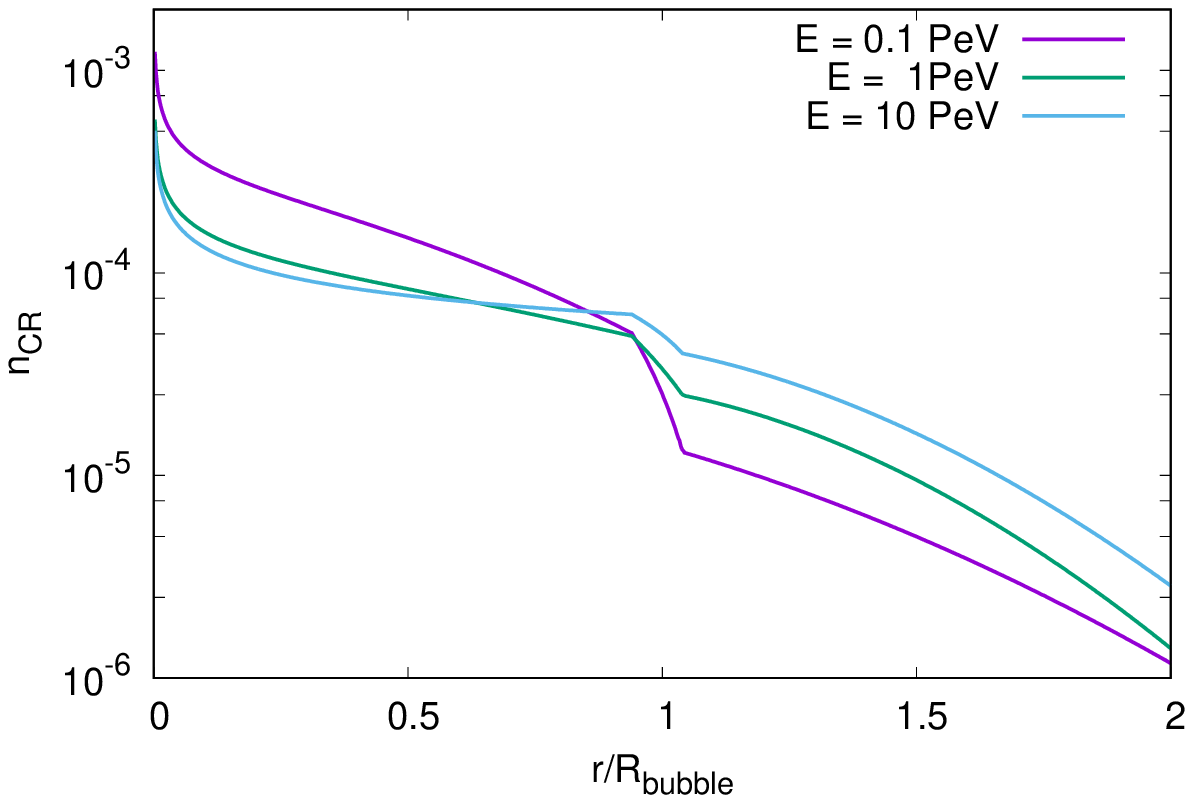}
\includegraphics[width=\columnwidth,angle=0]{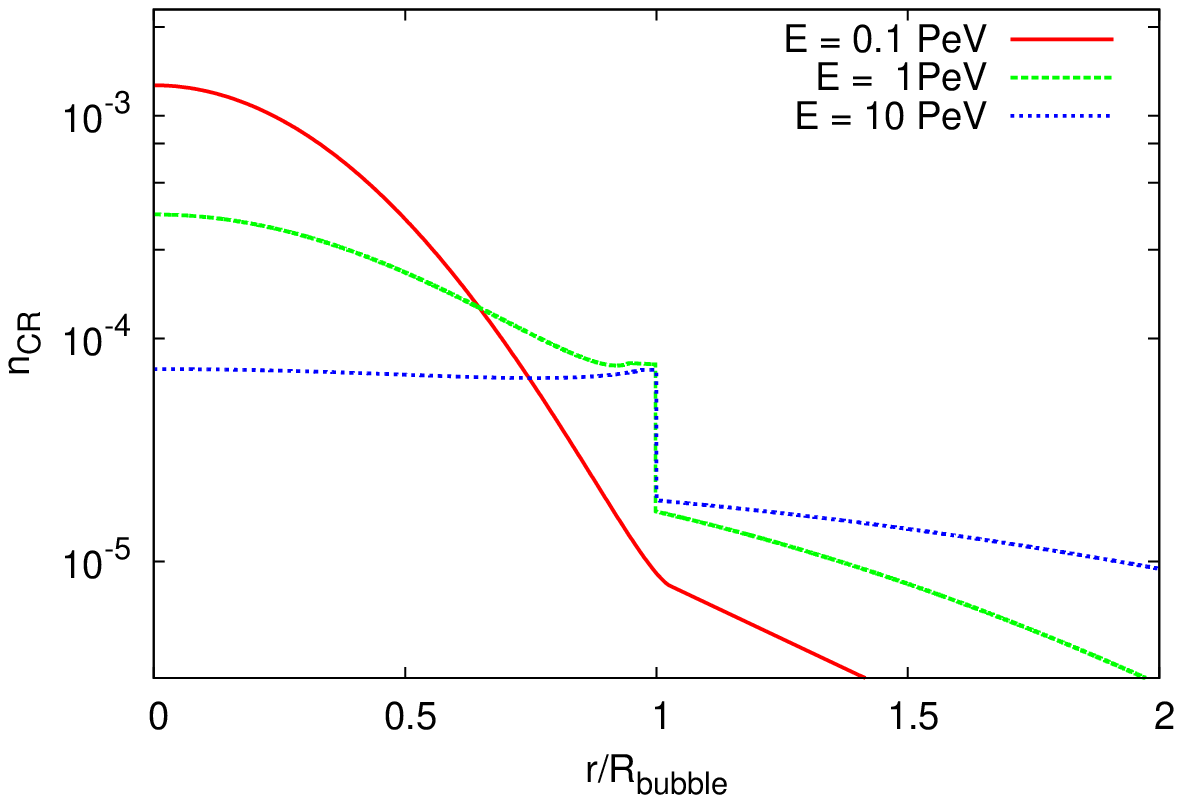}
\caption{Proton surface density of a continuous (left)
and a bursting (right)
source as function of distance in units of the bubble radius for 
three different energies.}
\label{fig:CR}
\end{figure*}

In Fig.~\ref{fig:CR} we show the normalised CR surface density of a
bursting and a continuous source (right) as function of distance in units
of the bubble  radius for three different energies~\citep{An17}. Let us 
discuss first the case of a bursting source shown in the left panel 
after the time $t=3000$\,yr: At low energies, $E\lsim 0.1$\,PeV, the
Gaussian diffusion front $\propto (2Dt)^{1/2}$ (with $D$ as the diffusion
coefficient inside the bubble) has not reached yet
the bubble wall. The small fraction of CRs which left already the
bubble feels the stronger magnetic field outside, leading to a slower decrease
of the CR density at $r>R$. Thus the CR density is described by a
quasi-Gaussian density profile with two effective
diffusion coefficients inside and outside the
bubble\footnote{\citet{An17} discusses fits of various analytical toy models to
    the CR density inside and outside  of the bubble.}.
In contrast,
at the highest energy considered, $E=10$\,PeV, CRs propagate inside
the bubble close to the ballistic regime, being then ``slowed'' down
by the increased diffusion in the bubble wall. Thus the bubble wall
acts as a kind of ``semi-permeable membrane'' increasing the CR density
inside the bubble. After a drop in the CR density, a quasi-Gaussian
tail is visible outside the bubble. Finally, for the intermediate
energy  $E=1$\,PeV, CRs diffuse also inside the bubble, and
 a quasi-Gaussian density profile with two effective
diffusion coefficients inside and outside the bubble is visible.
The main difference between the bursting and the continuous source
shown in the right panel is the absence of a characteristique energy
below which CRs have not yet reached the bubble wall. As a result,
the production of low-energy energy secondary photons is not suppressed.

\section{Loop~I and Local Superbubble}
%
We consider next the special case of our local neighbourhood. The Sun
is situated inside the Local Superbubble, a irregular formed volume of the
interstellar medium with an extremely low density and radius $\sim 100$\,pc.
\citet{1995A&A...294L..25E} noted the possibility of a
collision of the nearby Loop~I superbubble with the Local Bubble, forming a
wall of neutral and dense material in the interaction zone.
In this work, we will use the geometry sketched in their Fig.~5. 
In particular, we will assume that the Sun is close
to the wall, choosing as the smallest distance $d=25$\,pc. 
We assume that the interface between  the Local Bubble and Loop~I
has the density $n=20$/cm$^3$, while the remaining bubble wall of
Loop~I has the density  $n=10$/cm$^3$.

Compared to our calculations of CR propagation in our toy model,
Loop~I has a radius which is three times larger. Moreover, we
are interested in a source as Vela which is $\simeq 11.000$\,yr
old. From the scaling law  $L\propto (2Dt)^{1/2}$ we estimate that
we can use the results presented in Fig.~\ref{fig:CR} as a proxy
for the case  $t\simeq 11.000$\,yr and $R\simeq 100$\,pc appropriate for
the distance of Vela to the center of Loop~I. Alternatively, the CR source
may be older and the magnetic field strength in Loop~I higher than we
assumed.

\begin{figure}
\includegraphics[width=\columnwidth,angle=0]{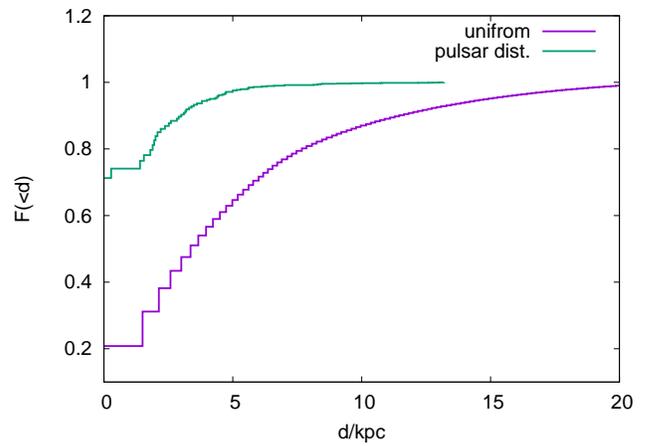}
\caption{The normalised cumulative flux of 100~source uniformly distributed
  in the Galactic disk (magenta line) and distributed as young
  pulsars (green line) according to \citet{Neronov:2012kz}.}
\label{fig:ex}
\end{figure}

Note that in a two-dimensional geometry as the Galactic disk,
sources inside a ring with distance $\rho$ to the observer contribute
as $1/\rho$ to the observed flux.
Therefore nearby sources contribute strongly even assuming a uniform
source distribution $n_s$. More explicitly, one can calculate the flux from
100~sources distributed uniformly in the Galactic disk. In Fig.~\ref{fig:ex},
we show the cumulative flux from sources within the distance $d$
normalised such that the total flux is one as a magneta line. While one
expects on average the nearest source at the distance $d\simeq 1.5$\,kpc,
the distance to Vela is  $d=0.3$\,kpc, leading to an enhancement of its
flux by a factor~25. This enhancement becomes stronger, if one takes into
account the known spatial distribution of pulsars and SNRs. For instance,
Figure~14 of \citet{Neronov:2012kz}
shows the location of known pulsars with age $< 30.000$yr (see also their
Table~III). Except for Vela, no other pulsar is known within 1\,kpc, and
many have distances as large as 3--5\,kpc. Using this distribution, one
obtains for the cumulative flux within the distance $d$  the green line.
We conclude  that Vela (or other sources in Loop~I
which are at an exceptional small distance compared to other known pulsars
or young SNRs can dominate the total flux from all Galactic sources. 
This implies also that the extragalactic flux from
normal galaxies gives a negligible contribution.

\section{Neutrino and photon fluxes}
%
We use the Monte Carlo generator QGSJET-II~\citep{Ostapchenko:2006nh,Ostapchenko:2010vb} 
to calculate the photon and neutrino secondary fluxes. 
We assume a mass fraction of 24\% of Helium in the target
gas and calculate the average intensity of the
secondaries as
\begin{align}
 I_i(E) & = \frac{c}{4\pi} \sum_{A\in\{1,4\}} \int_E^\infty{\rm d}E' \,
          \frac{{\rm d}\sigma_{\rm inel}^{pA\to i}(E',E)}{{\rm d}E}
\\ & \times
          \int{\rm d}^3x \,  \frac{n_p(E',\vec x) n_{\rm gas}^A(\vec x)}{d^2}\,,
\end{align}
where $\sigma_{\rm inel}^{pA}$ is the production cross section of secondaries
of type $i$ in interactions of protons on nuclei with mass number $A$,
$d$ denotes the distance from the Sun to the interaction
point $\vec x$, $n_p(E,\vec x)$ the differential number density
of CR protons and $n_{\rm gas}^A(\vec x)$  the density of protons
and Helium in the bubble wall of Loop~I, respectively. We use as
injection spectrum of CR protons
$dN/dE\propto E^{-2.2}\exp(-E/E_0)$ with $E_0=3\times 10^{15}$\,eV,
normalised such that the total energy emitted in CRs is
$E_{\rm CR}=2.5\times 10^{50}$\,erg. Then we
use the normalised CR surface density shown in Fig.~\ref{fig:CR} 
to obtain the relevant CR density inside the bubble wall.

\begin{figure}
\includegraphics[width=0.7\columnwidth,angle=270]{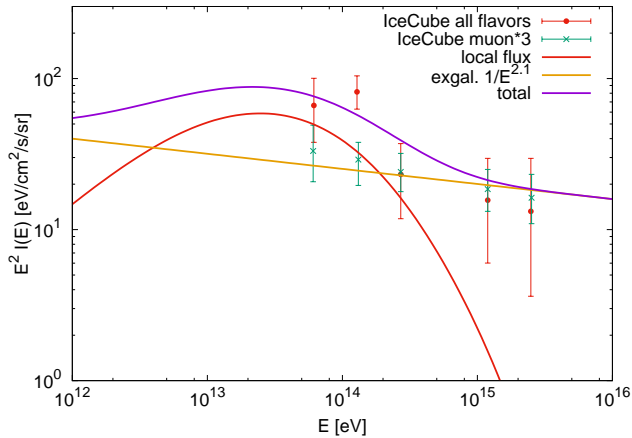}
\caption{Average neutrino intensity $E^2I(E)$ from Loop~I as function of
  energy $E$ is shown by a red line together with IceCube neutrino
  observations~\citep{Aartsen:2017mau}. 
  An  extragalactic component with spectral shape $1/E^{2.1}$
  is shown by a orange line and the sum of both components by a violet line.}
\label{fig:nu}
\end{figure}

In Fig.~\ref{fig:nu} we show by a red line the resulting intensity $I(E)$
multiplied by $E^2$ of neutrinos on Earth obtained in our model.  An
extragalactic component with spectral shape $1/E^{2.1}$ as fit to the muon
data is shown by an orange line. Finally, the total intensity as sum of
the two components is shown by a violet line.
While the neutrino intensity of the Galactic
component drops below $10^{14}$\,eV because CRs with energy lower than
$10^{15}$\,eV have not reached yet the bubble wall, it is suppressed at high
energies because of the assumed cutoff in the CR injection spectrum.
The combined neutrino
intensity of the Galactic and extragalactic contributions gives a good fit
of the experimental data~\citep{Aartsen:2017mau}.
In most concrete models for extragalactic
neutrinos, the predicted intensity is not a pure power law. For instance,
the neutrino intensity predicted in Ref.~\citep{Kachelriess:2017tvs} 
becomes steeper
than an $1/E^2$ power law below $10^{14}$\,eV, leading thus to a more
pronounced neutrino bump.
Since photon absorption plays for the small distances considered no role,
the corresponding photon flux is uniquely determined. An important constraint
on our model comes therefore from the limits on the diffuse gamma-ray flux
in the 100\,TeV--1\,PeV energy range~\citep{Ahlers:2013xia}, which were imposed
in particular by the KASCADE experiment from the non-observation of photon-like
events at those energies.  Note that this limit was reconsidered recently by
the KASCADE-Grande collaboration taking into account post-LHC hadronic models.

\begin{figure}
\includegraphics[width=\columnwidth,angle=0]{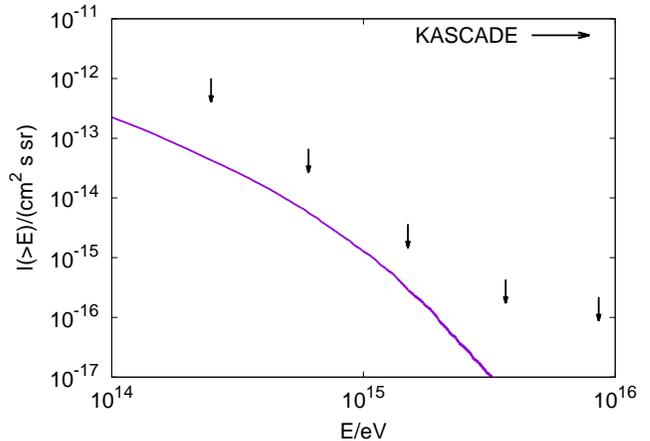}
\caption{Integral photon intensity $I(>E)$ from Loop~I
  as function of energy $E$ compared to upper $90\%$~C.L.
  derived from KASCADE data~\citep{Apel:2017ocm}.}
\label{fig:gam}
\end{figure}

In Fig.~\ref{fig:gam} we show that the integral photon intensity $I(>E)$
obtained in our model as function of energy $E$ obeys the upper
$90\%$~C.L. derived from KASCADE data~\citep{Apel:2017ocm}.
Note however that the predicted photon flux is only a factor few below the
KASCADE limit, which makes it detectable by future experiments.
Additionally, the arrival directions of photon-like events in the KASCADE
data could be used already now to constrain a possible flux enhancement
towards Loop~I.

Finally, let us comment on the deviation expected in our scenario
from an isotropic neutrino flux. The sky region in the direction of the
Loop~1 superbubble is approximately circular with radius $60^\circ$, centered
at at the Galactic coordinates $l=329.5^\circ$ and $b=17.5^\circ$. In the
6\,year IceCube data set, 15 out of 28 events  (54\,\%) with
$E> 100$\,TeV are located in this region, while one expect 8.6\, events
according to the IceCube exposure for an isotropic flux (or 31\%). Thus
the current dataset shows an excess in this region corresponding to 23\%
of the total number of events. Dedicated data analyses by the IceCube and the
ANTARES collaborations will have the potential to constrain or to favour this
scenario in the future.

\section{Conclusions}
%
The explanation of the IceCube data requires in addition to an extragalactic
component with a hard spectrum $1/E^\alpha$ and $\alpha=2.0-2.2$ a soft Galactic
component with the spectral slope $\alpha\simeq 2.5$
\citep{Neronov:2016bnp,Neronov:2015osa,Palladino:2017qda,Neronov:2018ibl}.
The bounds on the diffuse extragalactic gamma-rays suggest that such a
soft component has a Galactic origin. A Galactic neutrino component can be
decomposed into two contributions, depending on their angular distribution:
Neutrinos from sources in the Galactic plane  which are not nearby, and
neutrinos from local sources. The first component is strongly limited
by the recent bounds on the neutrino contribution from the Galactic
plane~\citep{Aartsen:2017ujz}.

In this {\em Letter\/} we studied the possibility that a nearby CR source
contributes to the astrophysical neutrino flux. In particular, we studied a
model where the CR source is located inside or nearby a superbubble, created
by previous supernovae. Both the magnetic field strength and the gas density
are enhanced in the wall of a superbubble. As a result, neutrinos are
preferentially produce in the wall of the superbubble. If the observer is
close to such a superbubble, neutrino events are distributed over a
considerable fraction of the sky. We applied then this mechanism to the case
of our local neighbourhood in the Galaxy. In particular, we considered the
Local and the Loop~I superbubbles which are interacting and have an
``interaction wall'' in between them. As an example for a young nearby CR
source may serve the Vela supernova which exploded 11000\,years ago.
The neutrino flux resulting in this model shown in Fig.~3 can be responsible
for a significant part of the IceCube neutrino flux at 
$\sim {\rm few} \times 100$\,TeV and below.
Other sources should provide only  one half of the
  flux predicted for Vela, cf.\ with Fig.~2.
A signature of this scenario are the correlation of the arrival direction
of Galactic astrophysical neutrinos with the matter distribution in the
walls of the Local and the Loop~I superbubbles. The contribution to the
neutrino signal from the former source would be rather isotropic, and thus
resemble the extragalactic component because we are sitting inside this
superbubble (but not at its center).

An important constraint on any Galactic neutrino model comes from the limits
on high-energy gamma-rays in the 100\,TeV--1\,PeV energy range. We showed that
the predicted photon flux in our model is only a factor few below the KASCADE
limits which makes a detection possible for future experiments. Such a
detection which would add additional angular information could confirm that
both neutrinos and gamma-rays are produced by cosmic ray interactions in the
wall between the Local and the Loop~I superbubbles.
However, the connection to a concrete cosmic ray source model has to be
additionally proven. For instance, secondary acceleration on the wall of
the Loop~I superbubble may be operating as an additional acceleration
mechanism~\citep{2001AstL...27..625B,Parizot:2004em,Ackermann:2011lfa}. 
With or without secondary acceleration, the
identification of a fraction of IceCube neutrinos as Galactic ones implies 
the existence
of a nearby cosmic ray  PeVatron. Vela as a young nearby SNR is a good
candidate for this source. In this case, the CR spectrum in the energy region
of the knee can be dominated by a single source such as Vela, as suggested
e.g.\ by \citet{Erlykin:1997bs}.  Such a scenario complements naturally
the ``local source'' proposal of \citet{Kachelriess:2015oua,Kachelriess:2017yzq} where a 2--3 Myr old source
dominates the CR energy spectrum in the 1--100\,TeV range. 

We conclude that our scenario---if confirmed by future observations---opens
up the  possibility to study a nearby PeVatron at work 
through multi-messenger observations with neutrinos, gamma-rays and cosmic
rays.

\acknowledgments
We would like to thank the anonymous referee for helpful suggestions.
This research was supported in part with computational resources at NTNU 
provided by NOTUR, \url{http://www.sigma2.no}.


\end{document}